\documentclass{iopart}
\begin{document}

\letter{Kurt Symanzik --- a stable fixed point beyond triviality}

\author{Frieder Kleefeld}

\address{Centro de F\'{\i}sica das Interac\c{c}\~{o}es Fundamentais (CFIF), Instituto Superior T\'{e}cnico, Av.\ Rovisco Pais, 1049-001 Lisboa, Portugal}
\eads{\mailto{kleefeld@cfif.ist.utl.pt}, URL: {\sf http:/$\!$/cfif.ist.utl.pt/$\sim$kleefeld/}}

\begin{abstract}
In 1970 Kurt Symanzik proposed a ``precarious'' $\phi^4$-theory with a negative quartic coupling constant as a valid candidate for an asymptotically free theory of strong interactions. Symanzik's deep insight in the non-trivial properties of this theory has been overruled since then by the Hermitian intuition of generations of scientists, who considered or consider this actually non-Hermitian highly important theory to be unstable. This short --- certainly controversial --- communication tries to shed some light on the historical and formalistic context of Symanzik's theory in order to sharpen our (quantum) intuition about non-perturbative theoretical physics between (non)triviality and asymptotic freedom.
\end{abstract}




\nosections
The fundamental laws of physics seem to follow a principle of beautiful simplicity, which challenges human imagination and intuition to the extreme. One example is the theory  of Quantum Electrodynamics (QED), which describes nature to an extreme of accuracy, yet virtually had to be declared dead by outstanding theoreticians due to its inherent problem of ``triviality'', i.e. the absence of interaction for infinite cut-off: 
\vspace{2mm}
\begin{indented} \item L.D.\ Landau states in 1959 \cite{Landau:1959}:
{\em ``$\ldots$ It was demonstrated by Pomeranchuk in a series of papers that, as the cut-off limit is increased, the physical interaction tends to zero, no matter how large the bare coupling constant is. $\ldots$ By now, the `nullification' of the theory is tacitly accepted even by theoretical physicists who profess to dispute it. $\ldots$  It therefore seems to me inopportune to attempt an improvement in the rigour of Pomeranchuk's proofs, especially as the brevity of life does not allow us the luxury of spending time on problems which will lead to no new results. $\ldots$''}.\\[1mm]
L.D.\ Faddeev writes on p.\ 82 in Ref.\ \cite{Faddeev:2001a}: {\em ``$\ldots$ In the USSR due to the zero charge result of L.\ Landau et al.\ for QED, field theory was virtually forbidden $\ldots$''}.\\[1mm] 
D.J.\ Gross writes on p.\ 92~f in Ref.\ \cite{Gross:2001cn}: {\em ``$\ldots$ the famous problem of zero charge, a startling result that implied for Landau that `weak coupling electrodynamics is a theory, which is, fundamentally, logically incomplete'. This problem occurs in any non-asymptotically-free theory. $\ldots$ Under the influence of Landau and Pomeranchuk, a generation of physicists was forbidden to work on field theory $\ldots$''}.\\[1mm]
Or to use the words of R.F.\ Streater in his ``Lost Causes in Theoretical Physics''~\cite{Streater:2005a} (downloaded on 14.6.2005): {\em ``$\ldots$ Although it is nearly proved that there are no solutions except the free or quasifree fields in four space-time dimensions to the construction of a scalar Wightman field via the lattice approximation, some people still hold out hope that a clever trick will be found to avoid the nearly proved fact that the only fixed point of the renormalisation procedure is a trivial field. $\ldots$ It is very demoralizing to be a research student working on a theory which will probably lead to a trivial theory, if it leads to anything. This is not to say that the techniques of constructive quantum field theory should not be studied. $\ldots$''}.
\end{indented}
\vspace{2mm}
Another example is the seemingly only \cite{Nobel:2004b,Nobel:2004a} candidate Quantum Chromodynamics (QCD) \cite{Muta:1987mz} for the ``non-trivial'' theory of strong interactions, which is yet lacking conclusive experimental evidence in what concerns the reality of gluons and its interface to experimentally verifiable asymptotic states, besides theoretical accumulating arguments in favour of scalar confinement (see for example Refs.\ \cite{Bicudo:2003ji,Kleefeld:2005a}) and well founded, yet unsettled concerns by a distinguished lattice QCD expert~\cite{Wilson:2004de}. 

Interestingly it has been shown for example by Bender and Jones \cite{Bender:1988ux} using the example of $\phi^4$-theory that triviality ($d>4$) and non-triviality ($d<4$) coexist in the infinitesimal vincinity of $d=4$ dimensions. Furthermore it has been argued by for example Consoli \etal \cite{Castorina:1990br,Branchina:1990ve,Branchina:1993rj} that $\lambda\,\phi^4$-theories, undergoing spontaneous symmetry breaking, are aymptotically free (see also Huang \cite{Huang:1993fw}). Moreover, Consoli and Stevenson provide a beautiful and unexpected outline of how the non-trivial phenomenon of spontaneous symmetry breaking takes place \cite{Consoli:1999ni}. 

It was as early as 1970, when Symanzik  \cite{Symanzik:1971a,Symanzik:1971vw,Symanzik:1973a} proposed an asymptotic free $\lambda\,\phi^4$-theory in the context of the restless and painful struggle towards a theory of strong interactions involving great scientists like for example --- among several others --- Nambu and Gell-Mann, which is beautifully described in Refs.\ \cite{Faddeev:2001a,Gross:2001cn,Nobel:2004a,Fritzsch:1984qe,'tHooft:1998pj,Shifman:2001cq,Bardeen:1972xk} and which led finally in 2004 to the well deserved Nobel Prize in Physics honoring the contribution of Gross, Politzer and Wilczek (obviously performed under the strong influence of e.g.\ Coleman, and in the presence of complementary or foregoing related research work by scientists like e.g.\ 't Hooft and Symanzik) {\em ``$\ldots$ for the discovery of asymptotic freedom in the theory of the strong interaction $\ldots$''}. To use the words of the ``The Nobel Prize in Physics 2004 --- Advanced Information'' \cite{Nobel:2004a}:
\vspace{2mm}
\begin{indented} \item {\em ``$\ldots$ experimentally verified scaling had a great impact on the physics community. The idea now was to understand how a physical theory could include scaling, and in 1970 Kurt Symanzik (d. 1983) argued that only a theory with a negative so-called $\beta$-function can imply scaling; the term `asymptotic freedom' was coined for this kind of theory.  $\ldots$ Symanzik himself discovered a quantum field theory with a negative $\beta$-function, namely one with a scalar field with a four-point interaction with a negative coupling strength. However a theory of this kind is not well-defined, since it does not have a stable particle spectrum. $\ldots$''}.
\end{indented} 
\vspace{2mm}
Yet Symanzik himself states \cite{Symanzik:1971a} in 1970 about his theory, which presumably Stevenson called ``precarious'' \cite{Stevenson:1985zy}, that {\em ``$\ldots$ we know of no reason why (the renormalized) $g$ should e.g. take positive rather than negative values $\ldots$''}.
Then he writes in a manuscript {\em ``A field theory with computable large-momenta behaviour''} received at 12.12.1972 and published at 13.1.1973 in Lettere Al Nuovo Cimento \cite{Symanzik:1973a}:
\vspace{2mm}
\begin{indented} \item {\em ``$\ldots$ In the current extensive discussion of $\varphi^4$ theory it is usually taken for granted that the renormalized coupling constant $g$ must be positive. As emphasized previously} \cite{Symanzik:1971a} {\em there is no known reason, axiomatic or otherwise, for $g>0$ to be required for a physically acceptable theory. The feeling that otherwise the theory cannot have a vacuum and particles of discrete mass is not rigorously founded as discussed near the end of this letter $\ldots$ One must not consider, however, the $g<0$ mode as an attempt to continue the $g>0$ one to negative $g$, which is certainly impossible analytically, but as an entirely different mode of $\varphi^4$ theory $\ldots$''}.
\end{indented}
\vspace{2mm}
In the last sentence Symanzik displays a very deep understanding of the underlying formalism required to construct correctly --- as done just most recently by Bender \etal \cite{Bender:2003ve,Bender:2005zz} (see also Ref.\ \cite{Acharya:2005bf}) ---  a valid asympotically free electrodynamics inspired by an (unfortunately incorrect) old argument \cite{Dyson:1952tj} of Dyson (see also Ref.\ \cite{Azam:2004dj}).

In addition to the sizable amount of ``traditional'' literature on Symanzik's precarious theory (see for example Refs.\ \cite{Symanzik:1971a,Symanzik:1971vw,Symanzik:1973a,Brandt:1975bc,Brandt:1975xn,Brandt:1979a,Kupiainen:1984we,Gawedzki:1985cf,Stevenson:1985zy,Stevenson:1986bq,Soto:1986xy,Bender:1988ux,Castorina:1990br,Branchina:1990ve,Ito:1991yc,Branchina:1993rj,Alhendi:1993yr,Huang:1993fw,Langfeld:1997rx,Li:2004bw}), there has recently developed a renewed highly topical interest in Symanzik's precarious theory in the context of the relatively new research field of PT-symmetric Quantum Theory \cite{PThistory1} (see also Refs.\ \cite{Bender:1998ke,Znojil:2001,Znojil:2004xw,PTworkshop:2003,PTworkshop:2004,PTworkshop:2005,Bender:2005a}). This field makes strong use of ideas developed in the context of quantum theories with indefinite metric (see for example Ref.\ \cite{Kleefeld:2004qs} and references therein)  and is some special case of a more general non-Hermitian Quantum Theory, a formulation of which by the author is in progress (see for example Refs. \cite{Kleefeld:2005a,Kleefeld:2004qs,Kleefeld:2003dx,Kleefeld:2003zj,Kleefeld:2002au} and references therein). 

We want to briefly mention here only the following important results concerning PT-symmetric Quantum Theory: The claim of Bender and Boettcher \cite{Bender:1998ke} in 1998 that the class of non-Hermitian, yet PT-symmetric Hamilton operators $H=p^2 + x^2 (ix)^\epsilon$ ($\epsilon>0$) has --- {\em due to its PT-symmetry} --- a {\em real} spectrum {\em bound} from below, was rigorously proven for a more general class of PT-symmetric Hamilton operators in 2001 mainly on the basis of Bethe-ansatz techniques \cite{Dorey:2001uw,Shin:2002vu,Caliceti:2005xu,Dorey:2004fk}. Furthermore it became clear that the construction of a meaningful scalar product for such Hamilton operators yielding a probability interpretation and being defined on contours in the complex $x$-plane yields essentially a non-Hermitian problem also for seemingly ``Hermitian--looking'' PT-symmetric Hamilton operators like the quantum-mechanical analogue of Symanzik's precarious $-\phi^4$-theory, i.e.~a $-x^4$-theory (see e.g.\ Refs.\ \cite{Znojil:2001,Bender:2002vv,Bender:2004ej,Mostafazadeh:2004tp,Mostafazadeh:2004mx})\footnote[6]{Note that a causal (local) Minkowski space-time implies non-Hermitian boundary conditions \cite{Kleefeld:2004qs}.}. As a final step PT-symmetric Quantum Mechanics has been extended most recently \cite{Bender:2004sa} to PT-symmetric Quantum Field Theory. This allows us now to transfer conclusions drawn in the context of the non-Hermitian $-x^4$-theory with sufficient care to Symanzik's precarious $-\phi^4$-theory.

On this basis it is interesting to recall the immediate reaction of the 2004 Nobel Prize winners to the foregoing ideas of Symanzik:
\vspace{2mm}
 \begin{indented} \item D.J.\ Gross and F.\ Wilczek write in their famous manuscript {\em ``Ultraviolet behavior of non-Abelian gauge theories''} \cite{Gross:1973id}, which has been received on 27.4.1973 and published on 25.6.1973 in Physical Review Letters: 
{\em ``$\ldots$ K.\ Symanzik (to be published) has recently suggested that one consider a $\lambda\varphi^4$ theory with a negative $\lambda$ to achieve UV stability at $\lambda=0$. However, one can show, using the renormalization-group equations, that in such theory the ground-state energy is unbounded from below (S.\ Coleman, private communication) $\ldots$''}.\\[1mm]
H.D.\ Politzer states in his famous paper {\em ``Reliable perturbative results for strong interactions?''}~\cite{Politzer:1973fx} received on 3.5.1973 and published also on 25.6.1973 in Physical Review Letters: {\em ``$\ldots$ $\lambda\varphi^4$ theory with $\lambda<0$ is ultraviolet stable (Ref.\ [K.\ Symanzik, DESY Report No.\ 72/73, 1972]) and hence infrared unstable but cannot be physically interpreted in perturbation theory. Using the computations of [S.~Coleman and E.~Weinberg, $\ldots$], for $\lambda<0$ `improved' perturbation theory is arbitrarily good for large field strengths. In particular, the potential whose minimum determines the vacuum decreases without bound for large field. $\ldots$''}.\\[1mm]
Later, in 1975, D.J.\ Gross claims on p.\ 186 ff in Ref.\ \cite{Gross:1975}: {\em ``$\ldots$ consider the most general theory involving only scalar fields, $\phi_i$ $\ldots$ where I have chosen the $\phi$'s to be real (a complex field can always be written in terms of its real and imaginary parts). $\dots$ I shall prove (following S.\ Coleman): Theorem. If a scalar theory has an interaction described by ${\cal L}_I=-\lambda_{ijkl}\phi_i\phi_j\phi_k\phi_l$ and is asymptotically free then the effective couplings $\bar{\lambda}_{ijkl}(t)$ must be such that the quartic form $\bar{\lambda}_{ijkl}(t)\phi_i\phi_j\phi_k\phi_l$ is non-negative as $t\rightarrow \infty$. Otherwise the vacuum energy is unbounded from below. Corollary. The coupling $\lambda$, of the theory ${\cal L}_I=-\lambda \phi^4$, must be positive. For if $\lambda<0$ then the theory is asymptotically free and $\bar{\lambda}(t)\stackrel{t\rightarrow\infty}{\longrightarrow}-1/t$ and thus $\bar{\lambda}\phi^4\rightarrow -\phi^4/t$ is not a positive form. $\ldots$''}
\end{indented}
\vspace{2mm}
 The statements of Gross, Wilczek, and Politzer (which are reflected in the reasoning of Refs.\ \cite{Nobel:2004a,Nobel:2004b} and unfortunately shared by a great majority of contemporary scientists due to the way how theoretical physics is presently taught in textbooks), which were interestingly written {\em after} the publication of Symanzik's manuscript Ref.\ \cite{Symanzik:1973a}, made use of the here {\em not applicable} assumption of an underlying {\em Hermitian} quantum field theory and were obviously more guided by intuition rather than a rigorous proof. Remarkably, Gross, Wilczek, and Politzer stand here in a great tradition, as even Landau himself argued already as early as 1958 \cite{Landau:1958}:
\vspace{2mm}
 \begin{indented} \item {\em ``$\ldots$ $\ldots$ Negative values of $g_0$ (for which, in the limit $\Lambda\rightarrow \infty$, $g_c$ may not vanish) are in general inadmissible because no stationary states of a boson system exist for $g_0<0$. Indeed, for boson fields a classical limiting case exists in which each state may contain many particles. For $g_0<0$ the energy of the classical field $\varphi$, $\ldots$, is not positive definite and can decrease indefinitely with increase of the field amplitude $\varphi$. Physically this means that it should be energetically possible for an infinite number of particles to be created from vacuum. Thus the vacuum cannot exist for $g_0<0$. $\ldots$''}.
\end{indented}
\vspace{2mm}
 Had Landau left aside here just once his brilliant intuition and elaborated slightly on the non-Hermitian nature of the problem, he would have --- probably --- anticipated immeditately, what has been beautifully summarized by Bender \etal in 2001 \cite{Bender:2001qx}:
\vspace{2mm}
 \begin{indented} \item {\em ``$\ldots$ all of the eigenvalues of the Hamiltonian $H=\frac{1}{2}p^2 +\frac{1}{2}m^2 x^2 - g x^4 (g>0)$ are real. Even though the} [Rayleigh-Schr\"odinger perturbation] {\em series} [for the ground state energy of $H$] {\em is not Borel summable, the imaginary part of the ground-state energy is exactly zero due to the presence of the soliton $\ldots$. The same result applies to the non-Hermitian ${\cal PT}$-symmetric }~[$-g\phi^4$]~{\em quantum field theory $\ldots$. However, again one must be very careful about nonperturbative effects.   
 $\ldots$''}.
\end{indented}
\vspace{2mm}
Based on what is stated above it is hardly possible to agree with the following very euphoric assessment by Prof.~Lars~Brink of the Royal Swedish Academy of Sciences, who stated in the presentation speach for the Nobel Prize in Physics~2004~\cite{Nobel:2004b}:
\vspace{2mm}
\begin{indented} \item {\em ``$\ldots$ This year's Nobel Prize completes the picture that the work behind several earlier prizes initiated and as a result we now know the fundamental building blocks and we have a description of the four fundamental forces. $\ldots$ The theory of Gross, Politzer and Wilczek successfully describes the physics of quarks, the matter from which we are to a very large extent built. Since the discovery, further research has shown that these theories are unique. No other theories can account for the experimental picture and it is wonderful to know that Nature has chosen the only theory that we have found to be possible. $\ldots$''}.
\end{indented}
\vspace{2mm}
On the contrary, it was Kurt Symanzik who gave us not only the first, yet also a very feasible \cite{Kleefeld:2005a} example that Nature has significantly more candidates for a theory of strong interaction at its disposal than intuitively imagined.
\ack
This work is dedicated to the memory of Kurt Symanzik. It has been supported by the
{\em Funda\c{c}\~{a}o para a Ci\^{e}ncia e a Tecnologia} \/(FCT) of the {\em Minist\'{e}rio da Ci\^{e}ncia, (Tecnologia,) Inova\c{c}\~{a}o e Ensino Superior} \/of Portugal, under Grants no.\ PRAXIS XXI/BPD/20186/99, SFRH/BDP/9480/2002, POCTI/\-FNU/\-49555/\-2002, and POCTI/FP/FNU/50328/2003.

\end{document}